\documentclass[aps,prl,citeautoscript,reprint,superscriptaddress,floatfix,footinbib,showkeys]{revtex4-1}
\usepackage[utf8]{inputenc}
\usepackage{amsfonts,amssymb,amsmath}
\usepackage{graphicx}
\usepackage[colorlinks=true,linkcolor=black,urlcolor=black,filecolor=black,citecolor=black]{hyperref}

\newcommand{\mr}[1]{\mathrm{#1}}
\newcommand{\be}{\begin{equation}}
\newcommand{\ee}{\end{equation}}

\newcommand{\uv}{\mr{\mu V}}

\newcommand{\rt}{R_\mr{T}}

\newcommand{\vb}{V_\mathrm{b}}
\newcommand{\nng}{n_\mathrm{g}}
\newcommand{\vg}{V_\mathrm{g}}
\newcommand{\cg}{C_\mathrm{g}}

\newcommand{\ec}{E_\mathrm{c}}

\newcommand{\ag}{A_\mr{g}}

\begin{document}
\title{Frequency to power conversion by an electron turnstile}
\author{Marco Marín-Suárez}\email{marco.marinsuarez@aalto.fi}
\author{Joonas T. Peltonen}
\author{Dmitry S. Golubev}
\affiliation{Pico group, QTF Centre of Excellence, Department of Applied Physics, Aalto University, FI-000 76 Aalto, Finland}
\author{Jukka P. Pekola}
\affiliation{Pico group, QTF Centre of Excellence, Department of Applied Physics, Aalto University, FI-000 76 Aalto, Finland}
\affiliation{Moscow Institute of Physics and Technology, 141700 Dolgoprudny, Russia}


\maketitle

\begin{figure*}
\includegraphics[scale=0.9]{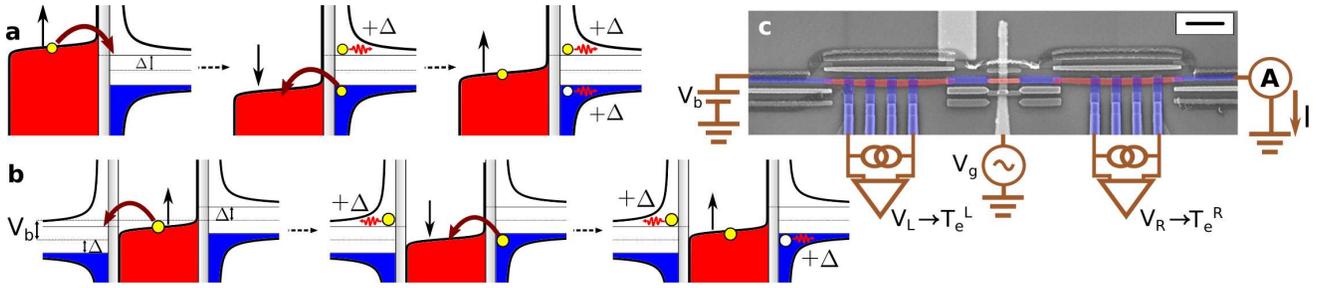}
\caption{\textbf{Single-electron turnstile for FPC.} (a) Sketch illustrating the operation at zero bias  exemplified in a hybrid single-electron box. The chemical potential of the N electrode (left) is varied periodically. First, an electron (yellow dot) has enough energy to be injected into the S lead (right) at the gap edge from the island as an excitation. Later, the driving enables an electron to tunnel from the lead to the island while the previously injected excitation diffuses. Once this electron tunnels to the island, it leaves one excited state close to the gap edge in the lead. Total energy of $2\Delta$ is injected into the lead per cycle. In the case of a turnstile at zero bias, the operation is the same as here, but the tunnelling events occur stochastically through the two contacts. (b) Sketch illustrating the non-zero bias behaviour in FPC. As opposed to the zero bias case, the excitations are created in both leads, giving again a total injected energy of $2\Delta$ per cycle distributed equally to the two leads. (c) Experimental setup for measuring the injected power in the turnstile operation  together with a coloured scanning electron micrograph. Light red refers to normal-metal and blue to superconductor. Scale bar is $1\,\mr{\mu m}$.}
\label{f1}
\end{figure*}

\textbf{Direct frequency to power conversion (FPC), to be presented here, links both quantities through a known energy, like single-electron transport relates an operation frequency $f$ to the emitted current $I$ through the electron charge $e$ as $I=ef$~\cite{Geerligs1990,Kouwenhoven1991,Pekola2007,Pekola2013,Kaestner2015,Giblin2020}. FPC is a natural candidate for a power standard resorting to the most basic definition of the watt-- energy, which is traceable to Planck's constant $h$, emitted per unit of time. This time is in turn traceable to the unperturbed ground state hyperfine transition frequency of the caesium 133 atom $\Delta\nu_\mr{Cs}$; hence, FPC comprises a simple and elegant way to realize the watt~\cite{SI}. In this spirit, single-photon emission and detection at known rates have been proposed and experimented as radiometric standard~\cite{Zwinkels2010,Chunnilall2014,Vaigu2017,Rodiek2017,Lemieux2019,Saha2020,Zhu2020,Tomm2021}. However, nowadays power standards are only traceable to electrical units, i.e., volt and ohm~\cite{Palafox2007,Waltrip2009,SI,Waltrip2021}. In this letter, we demonstrate the feasibility of an alternative proposal based on solid-state direct FPC using a SINIS (S = superconductor, N = normal metal, I = insulator) single-electron transistor (SET) accurately injecting $N$ (integer) quasiparticles (qps) per cycle to both leads with discrete energies close to their superconducting gap $\Delta$, even at zero drain-source voltage. Furthermore, the bias voltage plays an important role in the distribution of the power among the two leads, allowing for an almost equal injection $N\Delta f$ to the two. We estimate that under appropriate conditions errors can be well below $1\%$.}

\begin{center}
\begin{figure*}[ht!]
\includegraphics[scale=0.74]{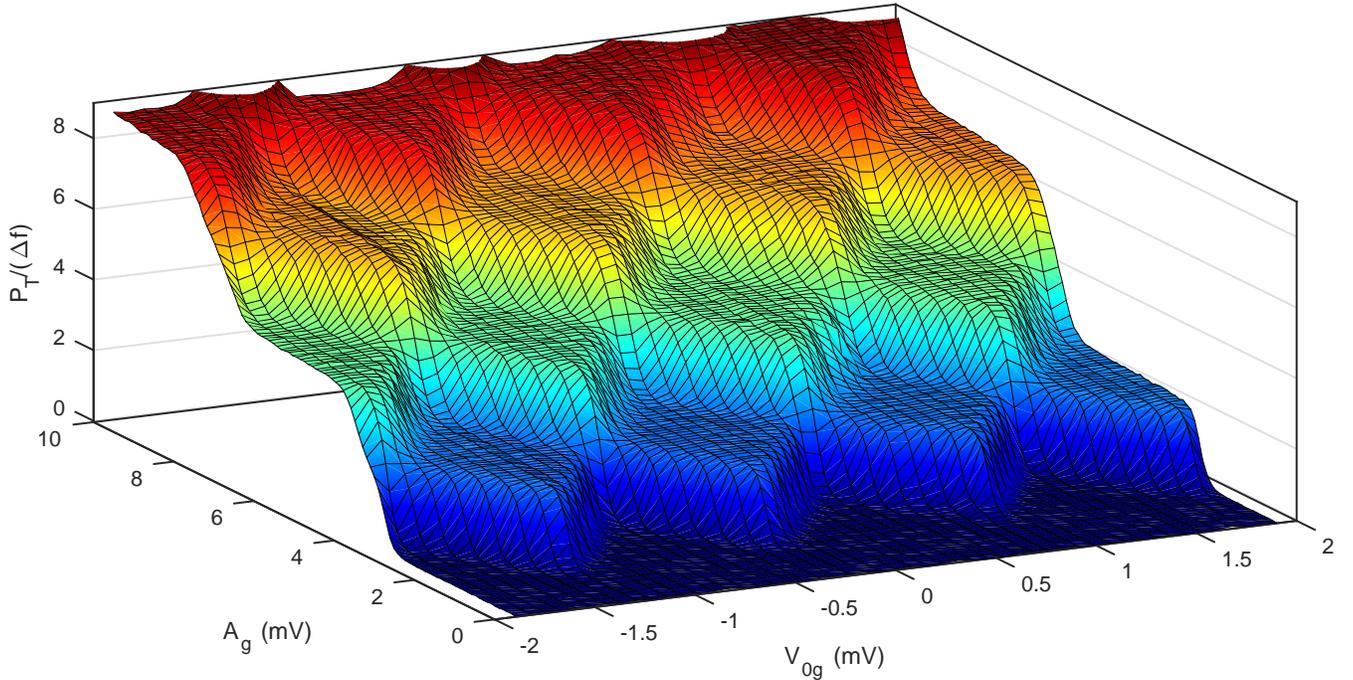}
\caption{\textbf{Power injection at zero bias.} Total injected power at $f=80\,\mr{MHz}$ and $\vb=0$ measured at $T_\mr{b}=130\,\mr{mK}$ as a function of the gate offset $V_\mr{0g}$ spanning several periods. The power is ideally given by $P=2N\Delta f$ with $N$ an integer. Here the gate amplitude $\ag$ range is such that four pumping plateaus become visible. It is evident that the injected power follows the diamond pattern, here even in the absence of average current through the device.}
\label{f2a}
\end{figure*}
\end{center}

The FPC relation can be understood based on a simplified picture of a driven NIS junction (Fig. \ref{f1}a) by looking at the qp injection dynamics. The key property here is the singularity of the superconducting density of states at energies $\pm\Delta$ as counted from the Fermi level. During the driving period where the chemical potential of N is periodically shifted, an electron tunnels into the superconductor with energy $\sim \Delta$ due to this diverging density of states. Later on, the driving provides enough energy for an electron to tunnel into the island breaking a Cooper pair and leaving an excitation in the superconductor, again close to the gap-edge. Thus, two tunnelling events per cycle occur; for larger driving amplitudes a higher even number $2N$ of tunnelling events is allowed. Within this picture, if a second superconducting lead is tunnel coupled to the island, the tunnelling events at bias voltage $\vb=0$ occur stochastically through either junction with probabilities proportional to each junction transparency. This whole operation results in total energy injection of nearly $2N\Delta$ in each cycle in the absence of net current. The case for a biased ($\vb\neq 0$) SINIS SET is depicted in Figure \ref{f1}b in which the qp injection events happen analogously to panel a. The only difference is that the bias defines a preferred direction for the charge transfer and therefore there will be $N$ tunnelling events per junction, i.e., the junction transparency does not play a key role anymore. The total energy injected is distributed almost equally to both leads but remains (nearly) unchanged. Time-averaged, the total injected energy current would be given by
\begin{equation}
P=2N\Delta f,
\label{e1a}
\end{equation}
and would exhibit a structure of plateaus similar to the charge current pumped through SINIS turnstiles~\cite{Pekola2007}, but now even at zero bias voltage. Accuracy of Eq. \eqref{e1a} can be tested in the FPC device depicted in Fig. \ref{f1}c (see Supplementary Section S1 for its characterization).

This device constitutes a turnstile for single electrons (see Supplementary Figure S1) when the island (light red short structure in Figure~\ref{f1}c) is periodically driven at frequency $f$ with a radio-frequency (rf) signal applied to a capacitively coupled, via the capacitance $C_\mr{g}$, gate electrode. At proper source-drain biases and driving amplitudes, an average charge current $I= Nef$ is obtained. This current is transported by $Nf$ qps injected per second to both tunnel-contacted leads (light blue short structure in Fig.~\ref{f1}c), respectively. The qps transport energy approximately without losses across the narrow leads~\cite{Peltonen2010,Knowles2012} to directly interfaced  normal-metal traps (light red long structures). These structures act as bolometers for measuring quantitatively the heat generated by the qp injection, accounting completely for the power~\cite{Ullom2000}. This heat can be determined within the conventional normal-metal electron-phonon interaction model~\cite{Giazotto2006} as $P=\Sigma\mathcal{V}\left(T_\mr{e}^5-T_\mr{b}^5\right)$. Here $\Sigma$ is the electron-phonon coupling constant of the material, $\mathcal{V}$ the trap volume, $T_\mr{e}$ its electron temperature and $T_\mr{b}$ the phonon bath temperature which usually can be taken as the cryostat temperature \cite{Dorozhkin1986}. The bolometer $\Sigma\mathcal{V}$ factor is calibrated \textit{in-situ} with an uncertainty $\sim 10\%$. See Supplementary Section S2 for details on the calibration of this detector. $T_\mr{e}$ is in turn obtained by current-biasing the superconducting tunnel probes (vertical blue structures in Fig. \ref{f1}c) contacted to the trap and measuring the corresponding voltage drop (see Supplementary Section S3 for the temperature calibration of these thermometers). The measured power can be compared to the expected FPC outcome.

\begin{figure*}[ht!]
\includegraphics[scale=0.8]{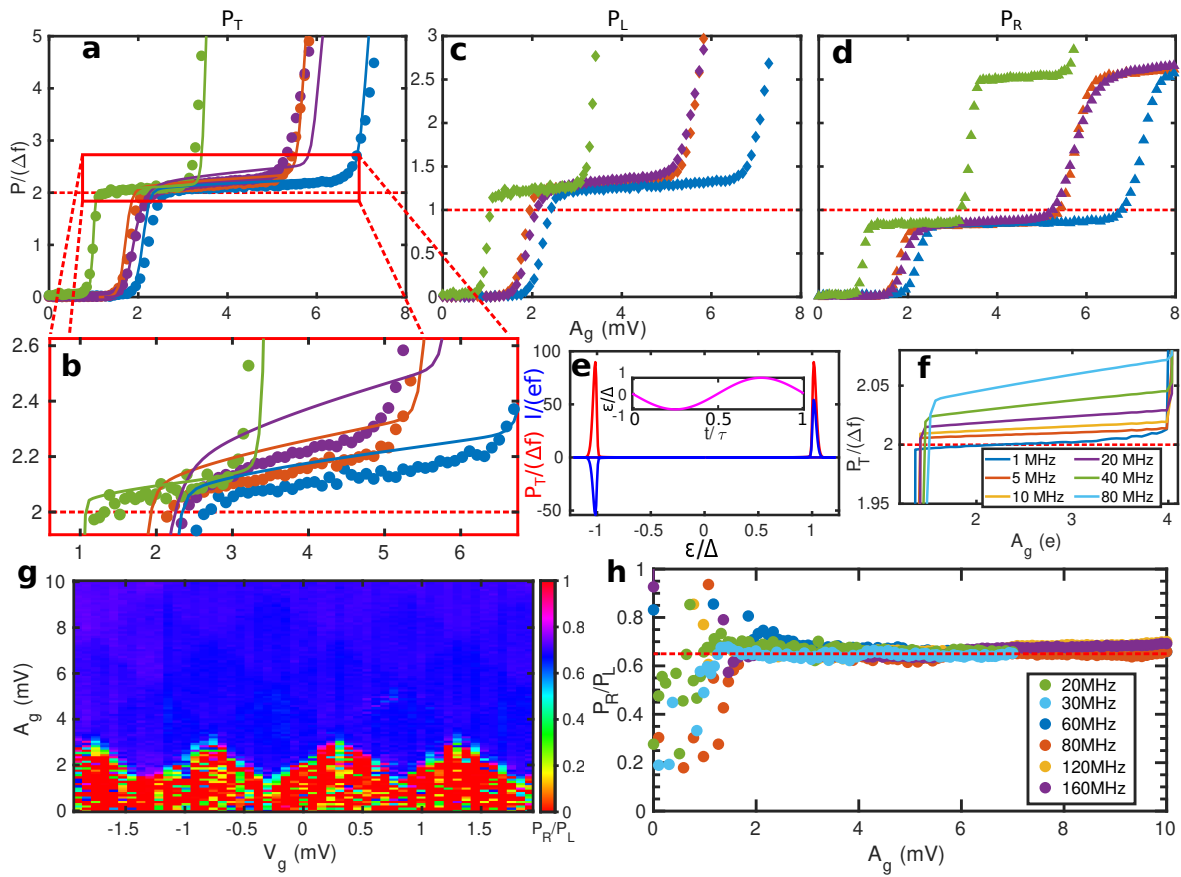}
\caption{\textbf{Power injection at zero bias.} (a) Power injected to the two leads in the absence of average current $\left(\vb=0\right)$ and at offset equivalent to $\nng=0.5$, showing that a power $\sim 2\Delta f$ for different driving frequencies is generated in the first plateau. Solid lines are calculated and data measured at $T_\mr{b}=117\,\mr{mK}$, colors correspond to the legend of panel (h). (b) Close-up of panel (a) to the first power plateau, showing how the calculated curves follow the trend of the data. (c) Power injected into the right lead and (d) into the left lead. (e) Simulation of instantaneous total injected power (red line) and current (blue line) as a function of the chemical potential difference for a jump into the island. The inset shows how this energy evolves within one sinusoidal driving cycle with a gate amplitude above the onset of the first plateau. Here, the measured device parameters are used and $f=5\,\mr{MHz}$, $T_\mr{S}^\mr{L}=T_\mr{S}^\mr{R}=T_\mr{N}=10\,\mr{mK}$. (f) Total injected power plateaus calculated for a turnstile with more optimized, but realistic, parameters (see the text) and with triangular gate driving. (g) Measured ratio between the injected power in the right and left leads as a function of the gate offset and amplitude, here the data are the same as in Fig. \ref{f2a}. Notice that this ratio is insensitive to the varied parameters and equal to the inverse of the normal-state junction resistance ratio. (h) Ratio between the injected power in the right and left leads at several driving frequencies and gate amplitudes, data from panels (c) and (d).}
\label{f2}
\end{figure*}

The main result of this work is shown in Fig. \ref{f2a} where the total injected power $P_\mr{T}$ (to the left $P_\mr{L}$ plus to the right lead $P_\mr{R}$) measured with the bolometers is shown. This was measured with a gate signal $\vg=V_\mr{0g}+\ag\sin{\left(2\pi ft\right)}$ at $f=80\,\mr{MHz}$ and $\vb=0$, hence no net charge (particle) current flows through the SET. Here, the DC gate voltage is swept over various gate periods revealing the $e$-periodic nature of the injected energy. Simultaneously, the signal amplitude is varied so that the gate-induced charge spans several charge stability regions in the Coulomb diamonds of the SET. The measured injected power exhibits plateaus of (approximately) constant value against $V_\mr{0g}$ and $A_\mr{g}$, following closely Eq.~\eqref{e1a} confirming the dynamics described in Figs. \ref{f1}a and \ref{f1}b. Therefore, we can assert that excitations are being injected close to the superconductor gap edge. The similarity of the plateaux pattern with that of Figure 2a in Ref. \onlinecite{Pekola2007} is evident and shows parallelism between the frequency to current conversion of single-electron transport and our proposal of frequency to power conversion, the latter being possible at wider bias ranges including $\vb =0$.

Further measurements of the power production at zero bias are shown in Fig. \ref{f2}. Panel \ref{f2}a shows the total injected power with $V_\mr{0g}$ at gate open position for a wide range of driving frequencies and confirms the results of Fig. \ref{f2a} at different injected powers. The transmission of the rf gate voltage depends on frequency giving different $\ag$ dependences of the otherwise similar power (and current) curves for different frequencies. Panel b shows that the total generated energy is close to ideal FPC. Indeed, the conversion errors range from $1.51\%$ to $6.32\%$ at low frequency $f=20\,\mr{MHz}$ and from $4.48\%$ to $14.26\%$ at higher frequency $f=160\,\mr{MHz}$. Calculations (solid lines) of the generated power resulting from a Markovian model (used also for DC characterization, See Supplementary Section S4) can reproduce the gate amplitude and frequency dependencies. Panels c and d exhibit the individual contributions to the power in the left and right lead, respectively. Notice that the two are not equal as explained above.

More insight into the dynamics of the zero bias operation is gained by simulating the instantaneous behaviour of the device. Figure~\ref{f2}e shows the calculated time dependent total injected power (Eq.~\eqref{power}, red curve) and current through one junction (Eq.~\eqref{current}, blue curve) plotted against the instantaneous energy change of an electron tunnelling into the island $\varepsilon \left(t\right)=2\ec\left(0.5-\nng\right)=-\left(2E_\mr{c}A_\mr{g}C_\mr{g}/e\right)\sin{\left(2\pi ft\right)}$. Here $\ec$ is the system charging energy and $\nng=\cg\vg/e$ is the normalized island excess charge induced by the gate voltage. A driving signal with offset equivalent to $\nng=0.5$, as used for most of the measurements at $\vb =0$, and period $\tau$ induces the evolution in $\varepsilon$ shown in the inset of Figure~\ref{f2}e for a driving amplitude at the start of the first plateau. It is clear that only electrons within a narrow energy band around $\Delta$ tunnel in and out of the island through the same junction, cancelling current and injecting an energy $\Delta$ per event as postulated. Naturally, the tunnelling events can happen through any junction. How often a given junction is involved depends on the relative transparencies to be explained below.

\begin{figure*}[ht!]
\includegraphics[scale=0.8]{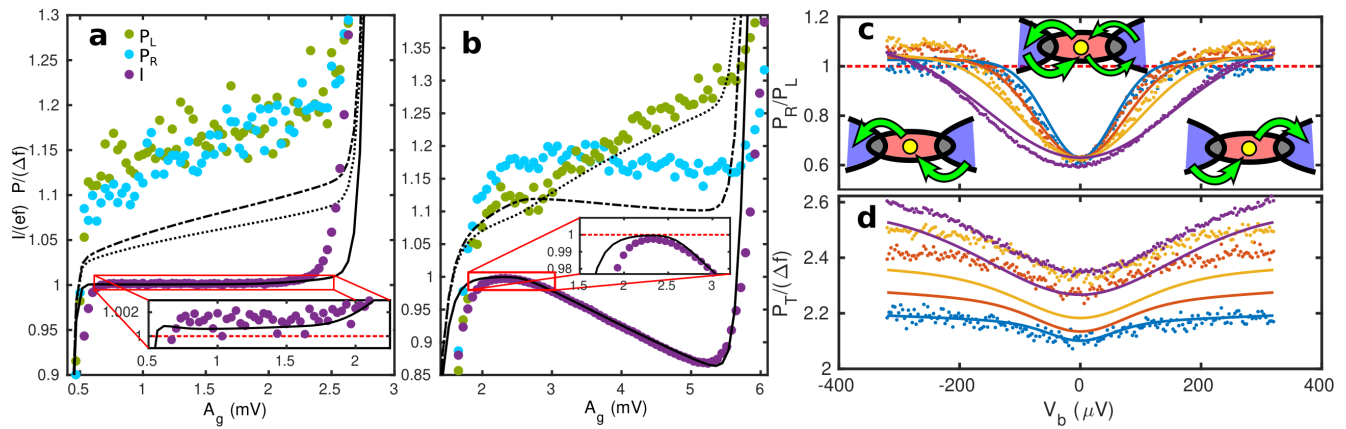}
\caption{\textbf{Power injection at non-zero bias.} Data measured at $T_\mr{b}=117\,\mr{mK}$ (a) Charge current (purple filled circles) and injected power to the left (green circles) and right (cyan circles) traps at $\vb=240\,\uv$ and $f=20\,\mr{MHz}$ against the gate amplitude. Black lines are simulations of charge current (solid), power injected into the left (dotted) and right (dash-dotted) leads, from the Markovian model. In this situation, the power is almost equally distributed to both leads. (b) As in (a) but for $\vb=160\,\uv$ and $f=60\,\mr{MHz}$, here the driving rate is comparable to the tunnelling rates. As a consequence the current and power injected to the right lead plateaus bend down and therefore the less transparent junction transmits less power. (c) Measured (dots) and calculated (solid lines) injected power ratios on the first plateau for driving frequencies $f=30,\, 60,\, 80,\, 160\,\mr{MHz}$ in blue, red, yellow and purple, respectively, as function of the bias. Two observations can be made, the ratio approaches 1 (i.e., one tunnelling event occurs per junction unidirectionally, see the cartoons) as the absolute value of bias increases (i.e., tunnelling rates increase) and it converges to $R_\mr{L}/R_\mr{R}$ at $\vb=0$ for all the frequencies. (d) As in (c) but for the total heat injected. The heat deviates from the value $P=2\Delta f$ when the driving frequency becomes comparable to the tunnelling rates.}
\label{f3}
\end{figure*}

Although the measured total injected powers shown in Figs. \ref{f2a} and \ref{f2}a are not exactly $2\Delta$ in the first plateau, we show in Fig. \ref{f2}f that a system with proper $\ec$, total normal-state junction resistance $\rt$ and Dynes parameter~\cite{Dynes1978} $\eta$ can inject a power closer to this value. In this panel the total injected power for a device with $\ec=\Delta=200\,\mathrm{\mu eV}$, $R_\mathrm{T}=200\,\mathrm{k\Omega}$ and $\eta=10^{-6}$ was calculated solving the same Markovian model used so far with low but achievable temperatures ($T_\mr{S}^\mr{L}=T_\mr{S}^\mr{R}=150\,\mr{mK},\,T_\mr{N}=10\,\mr{mK}$, for left and right lead and island, respectively). From there it is possible to see that when the driving is slow enough ($f=1\,\mathrm{MHz}$) the accuracy of the injected power is within $0.20\%$ and $0.65\%$ in the first plateau and between $1.27\%$ and $3.8\%$ for $f=80\,\mathrm{MHz}$. Therefore it is possible, in principle, to achieve a power injection of $2\Delta$ with small deviations given that $\eta$ is sufficiently small and temperatures are low. Under these conditions the ultimate limitation of accuracy would then be determined by the capability of measuring small powers bolometrically: at $1\,\mr{MHz}$ the power is $\sim 0.3\,\mr{fW}$ which exceeds the noise level in a standard setup by two orders of magnitude~\cite{Ronzani2018}. The bolometer noise is dominant over the shot noise of the generated energy flux caused by the stochastic nature of electron tunnelling, see Supplementary Section S5. Yet at these operation frequencies the Dynes parameter starts to play an important role making the power to be lower than Eq.~\eqref{e1a}. Additionally, we have verified that injection errors scale as $\left(\ec e^2\rt f/\Delta^2\right)^{2/3}$, thus higher power emission with smaller inaccuracy is possible for more transparent junctions. Yet in this argument we ignore the influence of Andreev tunnelling power injection~\cite{Rajauria2008}.

How the power is distributed to the two leads can be easily understood within the Markovian model (see details in Supplementary Section S6). The absence of a preferred direction of flow (i.e., the bias voltage) together with assuming the same qp temperature for both leads yield
\begin{equation}
\dfrac{P_\mathrm{R}}{P_\mathrm{L}}=\dfrac{R_\mr{L}}{R_\mr{R}}.
\label{e2}
\end{equation}
Where $R_\mr{L(R)}$ is the normal-state left (right) junction resistance. For the present device, we determined by the DC characterization $R_\mr{L}/R_\mr{R}=0.65$. The ratios shown in Figs. \ref{f2}g (calculated from data of Fig.~\ref{f2a}) and \ref{f2}h (from data in Figs. \ref{f2}c and \ref{f2}d) match this value, validating Eq.~\eqref{e2}. Thus, the power is distributed according to the ratio of junction transparencies. Notice the insensitivity of this quantity to gate parameters or temperature.

Next, we turn to the case $\vb\neq 0$. Figure \ref{f3} shows two paradigmatic cases in panels a and b (see Supplementary Section S7 for additional data). In the former, the driving is slow enough to enable fully synchronized tunnelling in the direction of bias, avoiding back-tunnelling (i.e., tunnelling against the bias). In the latter, the driving frequency is fast such that forward tunnelling is compromised and back-tunnelling may take place instead. As a result, the pumped current drops toward the end of the plateau.

In Fig. \ref{f3}a driving at $f=20\,\mr{MHz}$ and $\vb=240\,\uv$ was applied, avoiding the back-tunnelling. An accurate single-electron current (purple dots) is emitted whose behaviour is well described by our simulations (solid black line, see inset). Similarly to the zero bias case, Eq. \eqref{e1a} is closely followed (see green and cyan dots for left and right leads, respectively), however now the main difference lies in how the injected energy is distributed between the leads. The bias, and consequently a preferred direction of tunnelling enable an almost equal share in the power injection to both leads, following the dynamics described in Fig. \ref{f1}b. Both experiment and simulations show that now the distribution of the power has been inverted with respect to the zero bias case, i.e. $P_\mr{R}\gtrsim P_\mr{L}$. This can be explained by arguing that more energy needs to be provided for tunnelling through the more resistive junction, which is bound to happen under the present conditions before a back-tunnelling event can occur through the more transparent junction. In terms of Fig. \ref{f2}e the power peaks move to $|\varepsilon|\gtrsim\Delta$.

On the other hand, data in Fig. \ref{f3}b were measured with $f=60\,\mr{MHz}$ and $\vb=160\,\uv$. Towards high gate amplitudes a small current opposed to the bias flows; this behaviour is also well caught by our simulations. The ratio $P_\mr{R}/P_\mr{L}$ inverts with respect to the zero bias case for gate amplitudes close to the current onset (see the inset); this behaviour is as in panel a. When higher gate amplitudes provide enough energy to activate tunnelling opposed to the biased direction, this ratio inverts again and now the situation is analogous to the zero bias case although a preferred direction with non-vanishing net current is present in this regime. The dynamics of Fig. \ref{f1}b do not hold anymore and two tunnelling events can occur through the same junction within one cycle. Naturally, this process is bound to happen more likely through the more transparent junction, therefore $P_\mr{L}$ exceeds $P_\mr{R}$ as is clear from the figure. It is also evident that, as the current approaches zero due to back-tunnelling processes the power ratio decreases. Still, $P_\mr{R}/P_\mr{L}\geq R_\mr{L}/R_\mr{R}$, as shown in $P_\mr{R}/P_\mr{L}$ from Figure \ref{f3}c, where ratios for powers measured at different frequencies at the plateau are presented as function of the applied bias. The ratio collapses to a single value for all frequencies at $\vb=0$, as expected. Cartoons in Fig. \ref{f3}c exemplify the processes in the two presented regimes: when the voltage bias tends to zero the two tunnelling events are uncorrelated as to in which junction they occur. Conversely, at high biases these tunnelling events occur unidirectionally, one per junction. This is also shown in Supplementary Figure S8. For $f=30\,\mr{MHz}$ (blue dots), i.e., a driving slow enough to avoid back-tunnelling, relatively low biases give already $P_\mr{R}/P_\mr{L}\approx 1$, as expected. For higher frequencies, higher biases are needed to drive the ratio closer to $1$ (compare Supplementary Figs. S5--S7) since $\varepsilon$ changes at a rate comparable to the tunnelling rates giving rise to back-tunnelling. This fast change of $\varepsilon$ compared to the tunnelling rates enables electrons with a wider distribution of energies to tunnel, resulting in broader peaks than those shown in Figure \ref{f2}e, see Supplementary Figure S9. This gives as a result the pattern of Figure \ref{f3}d: for higher frequencies excitations with energies progressively higher than $\Delta$ are injected into the leads. In general, lower frequencies will give a more bias insensitive power injection. This behaviour is accurately captured by a Markovian model as seen from the solid lines in Figs. \ref{f3}c and \ref{f3}d.

As to further realizations and applications, FPC could also be realized by replacing the superconducting leads with quantum dots having a $\delta$-like singular density of states around tunable energy levels~\cite{Prance2009}. This tunability gives the possibility to modify at will the injected energy and might allow for increasing the conversion yield. Furthermore, the narrow highly peaked density of states around the energy level enhances the selectivity of the tunnelling events increasing the accuracy of the injected power. Importantly, the precision of the injection rate is ensured by the dot detuning from the Fermi level and the island finite charging energy. In addition to the present demonstration, FPC might find applications in nanoscale thermodynamics as a heat pump with no net particle flow~\cite{Rey2007,Arrachea2007,Kafanov2009,Hussein2016} as well as enabling a careful study of the dynamics of superconducting excitations because of the high injection control in our realization, unlike in recent injection demonstrations~\cite{Alegria2021}.

In summary, we have shown that a periodically driven NIS junction can be used as a synchronized power injector even in the absence of current. This is due to the singularity of the density of states in the superconducting lead which enables a high electron energy selectivity. This energy is then measured by a normal metal bolometer trapping the excitations. This, added to the possibility of injecting qps at a precisely known rate determined by the driving frequency, allows us to assert that a total power of $2N\Delta f$ is generated. The highly controlled injection allows one to measure a power as a known energy $\left(2N\Delta\right)$ released at a given repetition rate $\left(f\right)$ providing a natural realization of the unit of power. This rationale is completely in line with the ampere \textit{mise en practique} based on single-electron transport \cite{SI}. Our implementation has the advantage of being a real ``on-demand'' precise energy source, unlike single-photon sources where emission efficiencies do not exceed $50\%$~\cite{Tomm2021}. Because of the implemented SINIS geometry, the superconducting gap can be accurately measured \textit{in situ} by standard tunnel spectroscopy (as done here) or by determining the critical temperature. Notice that this is not possible in the system depicted in Fig.~\ref{f1}a. This necessity to measure $\Delta$ independently is the main difference between the proposed FPC and frequency to current conversion. While the electron charge is defined without uncertainty according to the 2019 redefinition of the SI base units, we estimate the uncertainty of our gap measurement to be $< 1\%$ (see inset in Fig. S1a and discussion). This defines the uncertainty of FPC because the one in frequency is negligible. Further steps of optimization, for instance in driving waveforms, are needed to achieve higher accuracy in power transfer. Although lower injection rates allow for more accurate conversion the detection method sets a low bound for the power generation. Most importantly, suitable device parameters and environmental conditions set lower errors in power emission. Furthermore, we demonstrated that the injection of qps follows a stochastic Markovian model favouring injection through the more transparent junction when the device is unbiased. When properly biased, the same amount of power is distributed evenly to the two leads.

\section*{METHODS}

\subsection*{Fabrication}

The samples were fabricated on 4-inch silicon substrates covered by $300\,\mr{nm}$ thermal silicon oxide. Masks were defined using electron beam lithography (EBL, Vistec EBPG500+ operating at $100\,\mr{kV}$) and metallic layers deposited using multi-angle shadow evaporation in an electron-beam evaporator. Directly on top of the substrate, ground planes and gate electrodes were formed by deposition of a $2\,\mr{nm}$ titanium adhesion layer, $30\,\mr{nm}$ gold, and a further $2\,\mr{nm}$  Ti protection layer over a mask defined in a single layer positive resist (Allresist AR-P 6200). This initial deposition is covered, after lift-off, by a $50\,\mr{nm}$ insulating $\mr{Al_2O_3}$ layer grown by atomic layer deposition(ALD). On top of this layer, a second EBL and metal evaporation process ($2\,\mr{nm}$ Ti followed by $30\,\mr{nm}$ AuPd) is carried out to shape bonding pads and coarse electrodes connecting to the transistor leads and two tunnel probes, the rest of the bonding pads and electrodes are patterned in the third and final step. After a second lift-off process, NIS transistor and probe junctions, clean NS contacts and remaining bonding pads and electrodes are formed by EBL patterning on a Ge-based hard mask process \cite{Pekola2013}. The mask is composed of a $\sim 400\,\mr{nm}$ P(MMA-MMA) copolymer layer, covered by $22\,\mr{nm}$ Ge also deposited by e-gun evaporation and a thin (approximately $50\,\mr{nm}$) layer of PMMA on top. After cleaving the wafer into smaller chips (typically $1\,\mr{cm}\times1\,\mr{cm}$), the pattern defined on the PMMA resist is transferred to the Ge layer by reactive ion etching (RIE) with $\mr{CF_4}$. Next, an undercut profile is created in the copolymer layer by oxygen plasma in the same RIE. Creation of tunnel junctions is done first by evaporating a $30\,\mr{nm}$ layer of Al at a substrate tilt angle of $-61.1^\circ$, resulting in a film that defines the finger-like superconducting probes. Right after deposition, this layer is oxidized \textit{in-situ} in the evaporator (static oxidation with typically $1.8\,\mr{mbar}$ for $1.5$ minutes). A subsequent deposition of $30\,\mr{nm}$ Cu at approximately $39.1^\circ$ tilt forms the normal-metal bolometers. Next, a second $30\,\mr{nm}$ layer of Al is evaporated at a tilt angle of $-32.5^\circ$ defining the transistor leads and the NS clean contacts. After a second oxidation (nominally $1.7\,\mr{mbar}$ $\mr{O_2}$ for one minute), a final $40\,\mr{nm}$ Cu film was deposited at normal incidence forming the N island of the SINIS transistor. After a final conventional lift-off step, a chip with an array of $3\times 3$ devices is cleaved to fit a custom-made chip carrier and electrically connected to it by Al wire bonds for measurements.

\subsection*{Measurements}

A custom-made plastic dilution refrigerator with base temperature of about $100\,\mr{mK}$ was used to carry out measurements. DC signals were applied through conventional cryogenic signal lines (resistive twisted pairs between room temperature and the $1\,\mr{K}$ flange, followed by at least $1\,\mr{m}$ Thermocoax cable as a microwave filter to the base temperature) connecting the bonded chip to a room temperature breakout box. Driving signals were transported to the gate by rf lines consisting of stainless steel coaxial cable down to $4.2\,\mr{K}$, a $20\,\mr{dB}$ attenuator in the liquid helium bath, followed by a feedthrough into the inner vacuum can of the cryostat. Inside the cryostat, the rf signal is carried by a continuous superconducting NbTi coaxial cable from the $1\,\mr{K}$ stage down to the sample carrier. At room temperature, an additional $40\,\mr{dB}$ attenuation is applied to the signal. Signals were realized by programmable voltage sources and function generators. Voltage and current amplification was achieved by room temperature low-noise amplifier (FEMTO Messtechnik GmbH, model DLPVA-100-F-D) and transimpedance amplifier (FEMTO Messtechnik GmbH, model DDPCA-300), respectively. The bath temperature is controlled by applying voltage to a heating resistor attached to the sample holder. The curves of the pumped current were typically repeated at least 10 times and averaged accordingly, neglecting those repetitions during which a random offset charge jump had occurred. Current amplifier offset was subtracted by comparing the pumping curves with their counterparts measured under source-drain bias of opposite polarity. The voltage drop curves across both bolometers were also repeated 15 times and averaged the same way as the current. After calibrating the bolometers' response against a previously calibrated ruthenium oxide thermometer (Scientific Instruments, Inc., model RO-600) attached to the cryostat sample carrier holder, the electronic temperature of the normal-metal trap is obtained by a linear fit to the response (see Supplementary Figure S3). Voltage amplifier offset is adjusted by comparing the response of the bolometer at equilibrium with its calibration curve and subtracting the difference.

\subsection*{System modelling}

The theoretical curves were obtained by calculating the current and power arising from the solution of a Markovian classical master equation on the island excess charge $n$
\begin{equation}
\dfrac{d}{dt}p\left(n,t\right)=\sum_{n\neq n'}\gamma_{n'n}p\left(n',t\right)-\gamma_{nn'}p\left(n,t\right).
\label{em1}
\end{equation}
Here $p\left(n,t\right)$ is the probability of the island to have $n$ excess charges at time $t$ and $\gamma_{nn'}$ is the total transition rate from the state $n$ to $n'$ which is directly related to the tunnelling rates through a NIS interface. The equation is solved in the steady state ($dp\left(n,t\right)/dt=0$) for the DC regime and with periodic conditions ($p\left(n,0\right)=p\left(n,\tau\right)$, with $\tau =1/f$) for the turnstile operation. The current through the left junction (L) is related to the occupation probabilities through
\begin{equation}
\begin{split}
I_\mr{L}&=e\sum_n{p\left(n\right)\left(\Gamma^\mr{L}_{n\rightarrow n+1}-\Gamma^\mr{L}_{n\rightarrow n-1}\right)}\\
&+2e\sum_n{p\left(n\right)\left(\Gamma^\mr{L}_{n\rightarrow n+2}-\Gamma^\mr{L}_{n\rightarrow n-2}\right)},
\end{split}
\label{current}
\end{equation}
where $\Gamma^\mr{L}_{n\rightarrow n\pm 1}$ denotes the single-electron elemental process rates and $\Gamma^\mr{L}_{n\rightarrow n\pm 2}$ second order Andreev process rates. The current can be averaged along one cycle as $\langle I_\mr{L}\rangle =1/\tau\int_0^\tau{dt I_\mr{L}}$.

The power injected to the transistor leads by stationary elementary events $\dot{Q}^\mr{R/L,S}_{n\rightarrow n\pm 1}$ gives the average injected power during one driving cycle as
\begin{equation}
\left\langle P_\mr{R/L}\right\rangle =\dfrac{1}{\tau}\int_0^\tau{dt\sum_n{p\left(n\right)\left(\dot{Q}^\mr{S,R/L}_{n\rightarrow n+1}+\dot{Q}^\mr{S,R/L}_{n\rightarrow n-1}\right)}}.
\label{power}
\end{equation}
In contrast to the current, the individual elementary tunnelling events contribute always additively to the power. For the DC case and for calculating the instantaneous power the integral is omitted. For obtaining accurate results comparable to experiments and because of the stiffness of the time periodic problem, an alternative numerical solution to Eq. \eqref{em1} based on propagation of the probability was carried out (see Supplementary Section S4). For further understanding of the instantaneous behavior of the quantities, Eq. \eqref{em1} was also solved at discrete cycle intervals using a variable order method.

\section*{ACKNOWLEDGEMENTS}

We acknowledge O. Maillet and E. T. Mannila for useful discussions. This research made use of Otaniemi Research Infrastructure for Micro and Nanotechnologies (OtaNano) and its Low Temperature Laboratory. This work is funded through Academy of Finland grant 312057, European Union's Horizon 2020 research and innovation programme under the European Research Council (ERC) programme (grant agreement 742559) and Russian Science Foundation (grant No. 20-62-46026).

\section*{AUTHOR CONTRIBUTIONS}
M.M.-S. made part of the fabrication, carried out the measurements, performed simulations and analysed the data with important input of J. P. P. and D. S. G. J.T.P. fabricated most part of the devices and prepared the measurement instruments. D. S. G. estimated the heat losses along the system. The primal idea was conceived by M.M.-S. and J.P.P. The manuscript was prepared by M.M-S. with important input from J.P.P, J.T.P and D.S.G.

\section*{COMPETING INTERESTS}

The authors declare no competing interests.

\section{DATA AVAILABILITY}

Data supporting the manuscript and supplementary Figures as well as further findings are available at \url{https://doi.org/10.5281/zenodo.5526576}.

\section{CODE AVAILABILITY}

The codes for generating the measured manuscript and supplementary Figures are available at \url{https://doi.org/10.5281/zenodo.5526576}. Algorithms for generating calculated curves are available from the corresponding author upon reasonable request.

\end{document}


\title{Supporting Information\\ Frequency to power conversion by an electron turnstile}
\author{Marco Marín-Suárez}\email{marco.marinsuarez@aalto.fi}
\author{Joonas T. Peltonen}
\author{Dmitry S. Golubev}
\affiliation{Pico group, QTF Centre of Excellence, Department of Applied Physics, Aalto University, FI-000 76 Aalto, Finland}
\author{Jukka P. Pekola}
\affiliation{Pico group, QTF Centre of Excellence, Department of Applied Physics, Aalto University, FI-000 76 Aalto, Finland}
\affiliation{Moscow Institute of Physics and Technology, 141700 Dolgoprudny, Russia}

\maketitle
\section{Device characterization and turnstile demonstration}

\begin{figure}[ht!]
\includegraphics[scale=0.95]{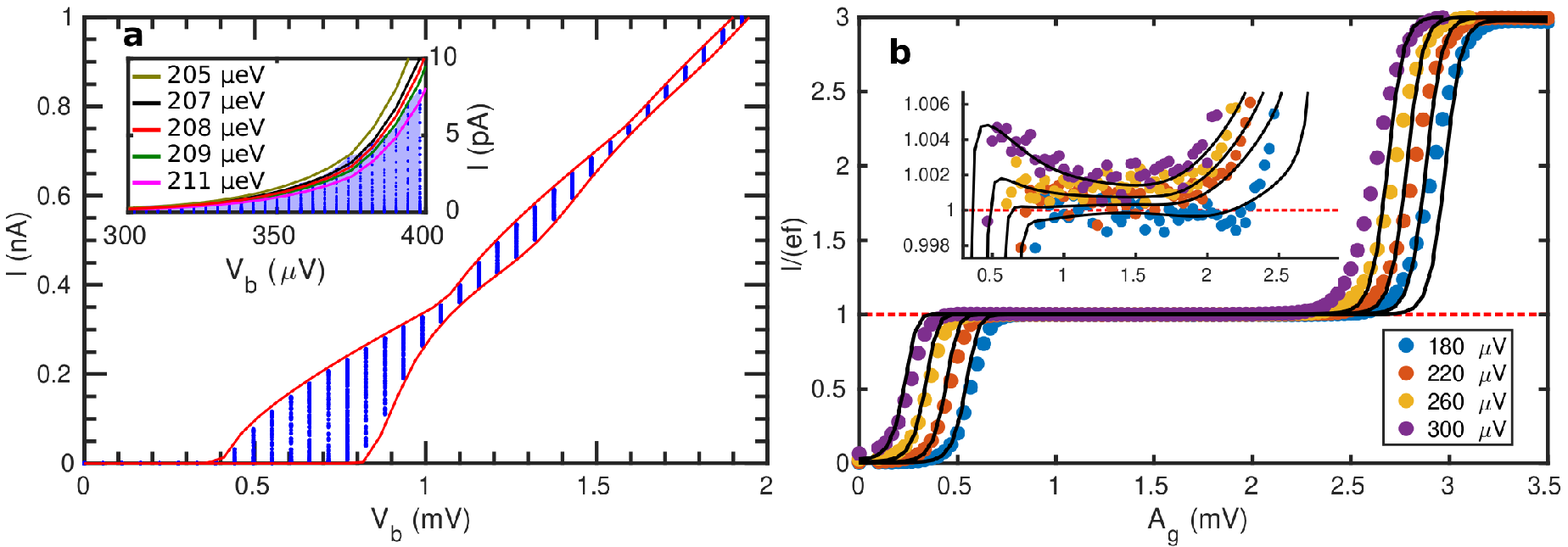}
\caption{\textbf{Device characterization and turnstile behaviour.} (a) DC IV curves for the measured sample. Red lines are simulations for the envelope bracketing experimental current at all the gate voltage values shown by the blue dots. (b) Typical turnstile measurements (dots) for the tested sample at $f=20\,\mr{MHz}$ ; black lines are simulations. Here the current is measured against the gate driving amplitude $\ag$, the dashed red line illustrates the ideal value of current $I=ef$. Inset: close-up of the plateaus around $I=ef$.}
\label{f1a}
\end{figure}

The device parameters of the SET used for FPC were estimated from a DC measurement in which the bias voltage $\vb$ is swept at the same time as a DC gate voltage. These measurements are shown in Fig. \ref{f1a}a as blue dots, red lines are the maximum and minimum transistor currents with respect to gate voltage calculated from a simple Markovian master equation on the discrete island charging events (described in Section \ref{S4}) using $\Delta=208\,\mr{\mu eV}$, charging energy $\ec=1.2\Delta$, total normal-state tunnel resistance $R_\mr{T}=1.36\,\mr{M\Omega}$, ratio between resistances $R_\mr{L}/R_\mr{R}=0.65$ and Dynes parameter \cite{Dynes1978} $\eta =2.75\times 10^{-4}$ as SET parameters, also a gate capacitance of $C_\mr{g}=0.12\,\mr{fF}$ was measured. Because of the importance of knowing $\Delta$ precisely in frequency to power conversion, we inspect how the tunnelling current near the threshold voltage changes with it in order to estimate an uncertainty. This comparison is shown in the inset, whose gate modulation range is highlighted by a blue shade for clarity. From this, we conclude that, in our case, the superconducting gap measurement has an uncertainty of $<1\%$ in the current realization.

In Figure \ref{f1a}b we demonstrate that this SET emits single-electron currents when voltage-biased and the gate is periodically driven. Here, a typical measurement is shown: the gate voltage is of the form $\vg=V_\mr{0g}+\ag\sin{\left(2\pi ft\right)}$, $V_\mr{0g}$ tuned at gate open position. One sees that the average measured current is around $I=ef$ for several gate amplitudes and $\vb$ values when $\vb\neq 0$. Solid black lines show results of simulations of the Markovian equation with the previously determined SET parameters. The inset shows that our calculations follow closely the measured current.

\section{Bolometer characterization}
\begin{figure}
\includegraphics[scale=0.95]{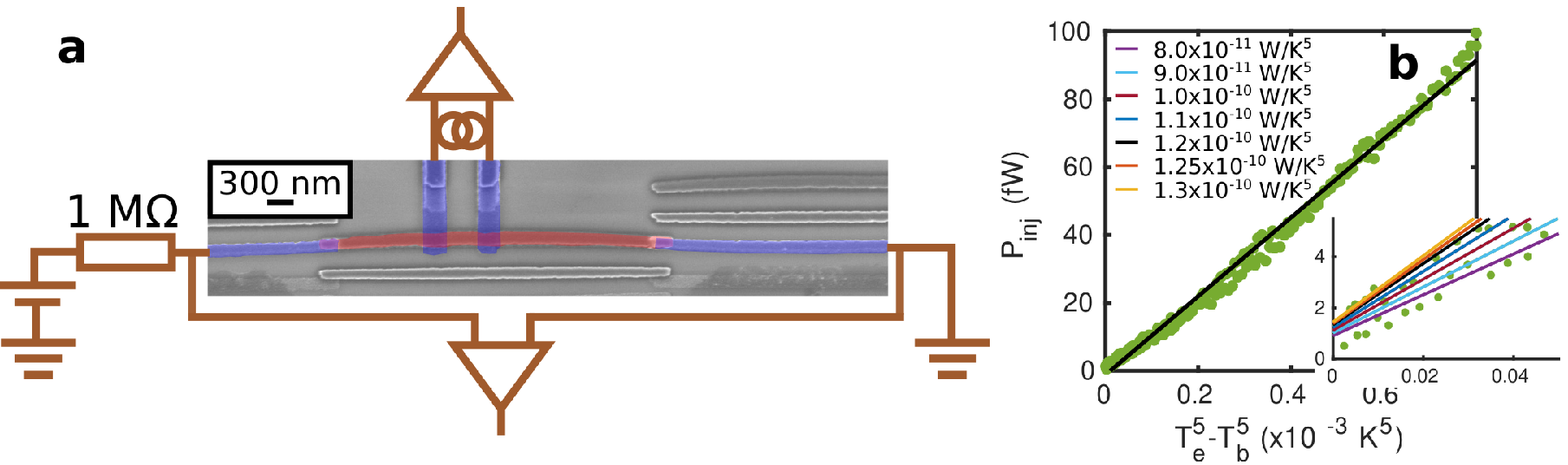}
\caption{\textbf{Bolometer characterization.} (a) Colored scanning electron micrograph of the device for measuring the electron-phonon coupling constant of the Cu strip (light red) along with the respective experimental setup. Light blue depicts superconducting Al. Dimensions of the normal-metal are nominally the same as in the measured device, which was evaporated during the same deposition on a different chip. (b) Measured injected power (green dots) against the difference of the 5th powers of the electron and bath temperatures. Black line is a linear fit to the data. The inset shows different linear fits with different values of $\Sigma\mathcal{V}$ in the power range of interest for our applications.}
\label{f1}
\end{figure}

In order to convert the measured temperature into the dissipated power $P$ through the conventional normal-metal electron-phonon interaction model,  e.g., $P=\Sigma\mathcal{V}\left(T_\mr{e}^5-T_\mr{b}^5\right)$, it is necessary to determine the quantity $\Sigma\mathcal{V}$. Since $\mathcal{V}$ is the volume of the normal-metal, it is reasonable to use the nominal values. However, since we have an overlap between the normal-metal and the superconductor in the clean SN contact of about $195\,\mr{nm}\times 112\,\mr{nm}$ we measure the value $\Sigma\mathcal{V}$ to avoid uncertainties due to the  qp relaxation in this region. This \textit{in-situ} calibration is important because we need a quantitative calibration for the actual device in the experiment. To achieve this, a Cu strip of nominally equal volume (see Fig. \ref{f1}a) was deposited on the same chip as the main samples over the same ground planes. Although the calibration of the bolometer was done on a different chip, its metal deposition was done at the same time as the measured sample to guarantee an identical behaviour of the Cu heat dissipation. Aluminium was evaporated to form SN contacts and NIS probes. The latter allow measuring the copper electronic temperature $T_\mr{e}$ as in the main sample, while the former allow the injection of a known current and measuring accurately the voltage drop in the normal-metal due to this current.

In Fig. \ref{f1}a a known current $I$ is generated by applying a voltage to a $1\,\mr{M\Omega}$ resistor, simultaneously the voltage drop $V$ in the normal metal is measured in a four-probe geometry. With this setup, the dissipated power is easily determined as Joule heating $P=IV$ which by continuity is also given by $P=\Sigma\mathcal{V}\left(T_\mr{e}^5-T_\mr{b}^5\right)$. Since we also have experimental access to $T_\mr{e}$ and bath temperature $T_\mr{b}$ it is possible to plot the power against $T_\mr{e}^5-T_\mr{b}^5$ as the green dots of Fig. \ref{f1}b, notice that the dependence is approximately linear confirming the exponent. The plotted data are well caught by a linear fitting (black line in Fig. \ref{f1}b) with a slope $\Sigma\mathcal{V}=1.2\times 10^{-10}\,\mr{WK^{-5}}$. By comparing the data with different slopes, an uncertainty interval for $\Sigma\mathcal{V}$ can be found. We conclude that $\Sigma\mathcal{V}=113\pm 13\,\mr{pWK^{-5}}$. We have decided to use the slope of the black line for plotting data merely for presentation purposes, since it is within the estimated uncertainty. Data converted to power match the Markovian model within the error interval. The present calibration shows that the slope for low powers is slightly smaller than for larger ones. In future experiments this needs to be studied further. The actual FPC measurements are performed in the power range $\lesssim 5\,\mr{fW}$.

\section{Bolometer response calibration}

\begin{figure}
\includegraphics[scale=1]{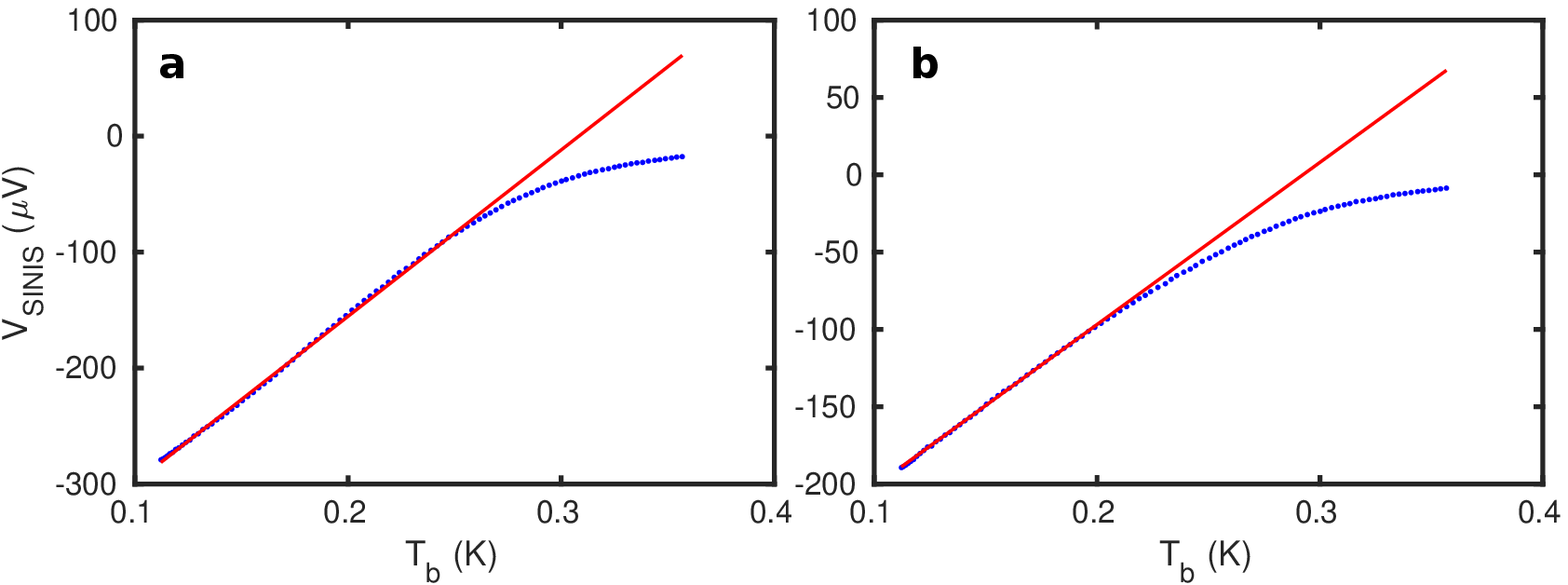}
\caption{\textbf{Bolometer response calibration.} (a) Response of the left lead bolometer (blue dots) to a bath temperature $T_\mr{b}$ and a fit to the linear region of the thermometer (red line). (b) As in (a) for the right lead.}
\label{f2}
\end{figure}

The thermometer responses were calibrated by comparing the measured voltage drop through the probe SINIS junctions $V_\mr{SINIS}$ with the bath temperature $T_\mr{b}$ measured by a ruthenium oxide resistor (Scientific Instruments, Inc., model RO-600) previously calibrated against a Coulomb blockade thermometer. This procedure was done by carefully controlling that no external heat sources (other than the environment) heated the bolometers. Therefore, a turnstile gate amplitude $\ag=0$ was applied throughout the whole calibration procedure. The results of these measurements are shown in Figs. \ref{f2}a and \ref{f2}b for the bolometer in contact with the left and right leads, respectively, as blue dots. There is a clear linear region whose slope is given by the red lines. We carefully keep the bath temperature and the response temperature due to turnstile power injection within the linear region during the main measurements.

\section{NUMERICAL CALCULATIONS}\label{S4}

In steady state (for DC calculations) and since the charging events $n$ are discrete one can express Eq. (3) as the matrix equation \cite{Saira2013}
\begin{equation}
A\mathbf{p}=\mathbf{0},
\label{e1}
\end{equation}
where $p_i=p(n_i)$, $A_{ii}=-\sum_{j\neq i}\gamma_{ij}$ and $A_{ij}=\gamma_{ij}$ for $i\neq j$. Extending to two leads, as for an SET, the total transition rates are given by $\gamma_{nn'}=\Gamma_{n\rightarrow n'}^\mr{L}\left(\varepsilon\right)+\Gamma_{n\rightarrow n'}^\mr{R}\left(\varepsilon\right)$. Here, $\Gamma$ is the transition rate between individual charging states $n$ and $n'$, $\mr{L}$ denotes events through the left junction and $\mr{R}$ similarly stands for the right one. Finally, $\varepsilon$ is the energy change of the process given by
\begin{align}
\varepsilon^{\pm,\mr{L/R}}_{1e}\left(n,V\right)&=\mp 2E_\mr{c}\left(n-n_\mr{g}\pm 0.5\right)\pm eV_\mr{L/R} \label{e6}\\
\varepsilon^{\pm,\mr{L/R}}_{2e}\left(n,V\right)&=\mp 4E_\mr{c}\left(n-n_\mr{g}\pm 1\right)\pm 2eV_\mr{L/R}, \label{e7}
\end{align}
for single-electron $\left(1e\right)$ and two-electron $\left(2e\right)$ tunnelling, respectively. Here $V_\mr{L}=\kappa_\mr{L}V_\mr{b}$ and $V_\mr{R}=-\kappa_\mr{R}V_\mr{b}$, $\kappa_\mr{L/R}$ is the ratio between the junction capacitance and the total capacitance, $V_\mr{b}$ is the bias voltage applied between the leads of the transistor. In Eqs.~\eqref{e6} and \eqref{e7} $n$ is the initial island excess charge, $n_\mr{g}$ is the charge number induced by the gate voltage, $E_\mr{c}$ is the charging energy and $+\,(-)$ designates tunnelling to (from) the island.

The explicit expression for the rates depends on the specific system and on the transition processes taken into account. Considering transitions in NIS junctions and only single-electron and two-electron Andreev processes, we have
\begin{equation}
\Gamma^\mr{L/R}_{n\rightarrow n\pm 1}\left(\varepsilon\right)=\dfrac{1}{e^2\rt}\int dE{n_\mr{s}\left(E\right)\left(1-f_\mr{N}\left(E+\varepsilon\right)\right)f_\mr{S}\left(E\right)} \label{e8}
\end{equation}
for single-electron tunnelling, and
\begin{equation}
\begin{split}
&\Gamma^\mr{L/R}_{n\rightarrow n\pm 2}\left(\varepsilon\right)=\dfrac{\hbar\Delta^2}{16\pi e^4\rt^2\mathcal{N}}\int dEf_\mr{N}\left(E-\varepsilon/2\right)f_\mr{N}\left(-E-\varepsilon/2\right)\times\\
&\left|a\left(E+E_\mr{c}-i\delta/2\right)+a\left(-E+E_\mr{c}-i\delta/2\right)\right|^2 \label{e9}
\end{split}
\end{equation}
for Andreev tunnelling. Here, $\varepsilon$ is the corresponding energy change from Eqs.~\eqref{e6} or \eqref{e7}, respectively.

In Eqs.~\eqref{e8} and \eqref{e9}, $\Delta$ is the superconducting gap of the leads, $\rt$ is the junction normal-state tunnel resistance and $\mathcal{N}$ is the number of conduction channels which can be written as $A/A_{\mathrm{ch}}$ with $A$ being the junction area ($\approx 49\,\mr{nm}\times 63\,\mr{nm}$ for the measured device) and $A_{\mathrm{ch}}$ is the area of an individual channel (typically $30\,\mr{nm^2}$). The term $\delta$ takes into account the energy of the intermediate (single-electron tunnelling) state which has a finite lifetime \cite{Averin2008} and therefore can be estimated as $\hbar\sum_\pm{\Gamma_{n\rightarrow n\pm 1}}$. However, in our simulations and in the regime of interest the precise value does not have an impact on the calculated Andreev tunnelling rates~\cite{Maisi2014a}. We set a reasonable value of $\delta/\Delta =10^{-5}$. Additionally, $f_\mr{N}$ is the Fermi-Dirac distribution of the normal-metal island, $f_\mr{S}$ is that for the superconducting lead involved in the tunnelling event and $n_s$ is the superconducting quasiparticle density of states given by 

\begin{equation}
n_\mr{s}\left(E\right)=\left|\mathfrak{Re}\left(\dfrac{E/\Delta+i\eta}{\sqrt{\left(E/\Delta+i\eta\right)^2-1}}\right)\right|,
\label{e11}
\end{equation}
here, $\eta$ is the Dynes parameter which models subgap leaks~\cite{Dynes1978,Pekola2010}. Furthermore,
\begin{equation}
a\left(x\right)=\dfrac{1}{\sqrt{x^2-\Delta^2}}\ln\left(\dfrac{\Delta-x+\sqrt{x^2-\Delta^2}}{\Delta-x-\sqrt{x^2-\Delta^2}}\right).
\end{equation}

Once the probability vector $\mathbf{p}$ is obtained by solving Eq.~\eqref{e1}, the current through the SET can be calculated as $I=\mathbf{b}\cdot\mathbf{p}$ with $b_i=e\left(\Gamma_ {i\rightarrow i+1}^\mr{L}-\Gamma_{i\rightarrow i-1}^\mr{L}\right)+2e\left(\Gamma_ {i\rightarrow i+2}^\mr{L}-\Gamma_{i\rightarrow i-2}^\mr{L}\right)$. In order to get realistic results the temperature change of the island has to be taken into account. To do this, one calculates the power transferred in single-electron events to the normal island provided $\varepsilon$ is constant
\begin{equation}
\dot{Q}^{\mathrm{N},\mr{L/R}}_{n\rightarrow n\pm 1}\left(\varepsilon\right)=\dfrac{1}{e^2\rt}\int{dEEn_\mr{s}\left(E-\varepsilon\right)f_\mr{N}\left(E\right)\left(1-f_\mr{S}\left(E-\varepsilon\right)\right)},
\label{e13}
\end{equation}
where N refers to normal metal and $\varepsilon$ is again the related energy change cost from Eq. \eqref{e6}. The total power transferred to the island is calculated as $\dot{Q}=\mathbf{q}\cdot\mathbf{p}$ with $q_i=\dot{Q}^\mathrm{N}_{i\rightarrow i+1}+\dot{Q}^\mathrm{N}_{i\rightarrow i-1}$, where $\dot{Q}^\mathrm{N}_{i\rightarrow i\pm 1}=\dot{Q}^{\mathrm{N},\mr{R}}_{i\rightarrow i\pm 1}+\dot{Q}^{\mathrm{N},\mr{L}}_{i\rightarrow i\pm 1}$. The heat flow to the phonons in the N island is governed by $\dot{Q}_\mathrm{e-ph}=\mathcal{V}\Sigma\left(T_\mathrm{N}^5-T_\mathrm{b}^5\right)$ where $\mathcal{V}$ is the volume of the island ($\approx 930\,\mr{nm}\times 105\,\mr{nm}\times 40\,\mr{nm}$ for the measured device), $\Sigma$ is the electron-phonon coupling constant ($\approx 8.4\times 10^9\,\mr{WK^{-5}m^{-3}}$), $T_\mathrm{N}$ is the electron temperature of the island and $T_\mathrm{b}$ is the temperature of the phonon bath which is considered to be the same as the base temperature. Finally, the heat balance is $\dot{Q}_\mathrm{e-ph}=\dot{Q}$. This condition is used to solve for $T_\mathrm{N}$.

A similar method can be applied during the turnstile operation~\cite{Saira2013}. If the period of the driving is $\tau$ then it is valid to assume that the steady state probability satisfies $p\left(t\right)=p\left(t+\tau\right)$. In order to solve for the probability the cycle is discretized in $m$ intervals of length $\Delta t=\tau/m$, next the matrix $A\left(k\Delta t\right)=A_k$ is calculated for each interval as well as the $\mathbf{b}_k$ vector. If we now build the propagator
\begin{equation}
U\left(\tau\right)=\prod_{k=1}^m\exp{\left(\Delta tA_k\right)},
\end{equation}
then the initial probability is given by
\begin{equation}
U\left(\tau\right)\mathbf{p}(0)=\mathbf{p}(0),
\label{e2}
\end{equation}
since for a periodic driving $\mathbf{p}(0)=\mathbf{p}(\tau)$.

A more useful form of the propagator is given by
\begin{equation}
\tilde{U}\left(\tau\right)=\prod_{k=1}^m\exp{\left(\Delta t\tilde{A}_k\right)},
\end{equation}
where
\begin{equation}
\tilde{A}_k=
\begin{bmatrix}
A_k & \mathbf{0}\\ \mathbf{b}_k^\mr{T} & 0
\end{bmatrix}.
\label{e10}
\end{equation}

In Eq.~\eqref{e10} $\mathbf{0}$ is a vector of zeros with the appropriate dimensions. This way Eq. (4) can be reformulated in terms of the augmented rate matrix $\left(\tilde{A}_k\right)$ and a new probability vector $\mathbf{\tilde{p}}\left(t\right)=\left[\mathbf{p}\left(t\right)\quad\langle q\rangle\right]^\mr{T}$, where $\langle q\rangle$ is the average charge transferred during one cycle. The final propagator can be decomposed as
\begin{equation}
\tilde{U}\left(\tau\right)=
\begin{bmatrix}
U & \mathbf{0}\\ \mathbf{U}_\mr{b} & 0
\end{bmatrix}.
\label{e12}
\end{equation}
Then, Eq.~\eqref{e2} is reformulated in terms of this new propagator and the average charge is obtained as $\langle q\rangle=\mathbf{U}_\mr{b}\cdot \mathbf{p}(0)$. The average current pumped is $I=\langle q\rangle/\tau$. The energy injected to the left/right superconducting leads is calculated by replacing $\mathbf{b}$ by $\mathbf{q}^\mr{L/R}$ in Eq.~\eqref{e10} with $q^\mr{L/R}_i=\dot{Q}^\mr{S,L/R}_{i\rightarrow i+1}+\dot{Q}^\mr{S,L/R}_{i\rightarrow i-1}$, where in turn $\dot{Q}^\mr{S,L/R}_{i\rightarrow i\pm 1}=\dot{Q}^\mr{N,L/R}_{i\rightarrow i\pm 1}-\varepsilon^{\pm,\mr{L/R}}_{1e}\Gamma^\mr{L/R}_ {i\rightarrow i\pm 1}$. Furthermore, the probability vector is redefined as $\mathbf{\tilde{p}}\left(t\right)=\left[\mathbf{p}\left(t\right)\quad\langle E_\mr{L/R}\rangle\right]^\mr{T}$, where $\left\langle E_\mr{L/R}\right\rangle$ is the average injected energy to the left/right lead and is obtained as $\left\langle E_\mr{L/R}\right\rangle=\mathbf{U}^\mr{L/R}_\mr{q}\cdot\mathbf{p}\left(0\right)$, where $\mathbf{U}^\mr{L/R}_\mr{q}$ is retrieved from a decomposition similar to Eq.~\eqref{e12}. Finally, the average injected power to the left/right lead is given by $P_\mr{L/R}=\left\langle E_\mr{L/R}\right\rangle/\tau$.

\section{FPC and Bolometer noise comparison}

From analogy to the current shot noise in single-electron pumps~\cite{Maire2008} and assuming that it is much more probable to have extra tunnellings than missed tunnellings; we estimate that the signal-to-noise ratio in FPC follows
\begin{equation}
\mr{SNR}_\mr{FPC}\propto \sqrt{\dfrac{P_\mr{ideal}f\tau}{P-P_\mr{ideal}}}.
\label{e17}
\end{equation}

Where $P_\mr{ideal}$ is the power produced with no extra/missed tunnelling events, $f$ the operation frequency, $\tau$ the averaging time and $P$ the real (with tunnelling errors) produced power.

On the other hand, there is noise in the power detection by the bolometer. The noise power associated with the power dissipation is given by
\begin{equation}
S_P=2k_\mr{B}T^2G_\mr{th}.
\end{equation}
Here $k_\mr{B}$ is the Boltzmann constant, $T$ the bolometer temperature and $G_\mr{th}$ the linear response thermal conductance of the bolometer taken as $G_\mr{th}=5\Sigma\mathcal{V}T^4$. The bolometer power dissipation signal-to-noise ratio is the given by
\begin{equation}
\mr{SNR}_\mr{bolo}=\dfrac{P}{\sqrt{S_P/\tau}}=\dfrac{P}{T}\sqrt{\dfrac{\tau}{2k_\mr{B}G_\mr{th}}}.
\label{e18}
\end{equation}
$P$ is also the power produced by FPC. At low frequencies ($f\ll G_\mr{th}T/(10\Delta)$) we can assume that $T\approx T_\mr{b}$.

\begin{figure}
\includegraphics[scale=0.8]{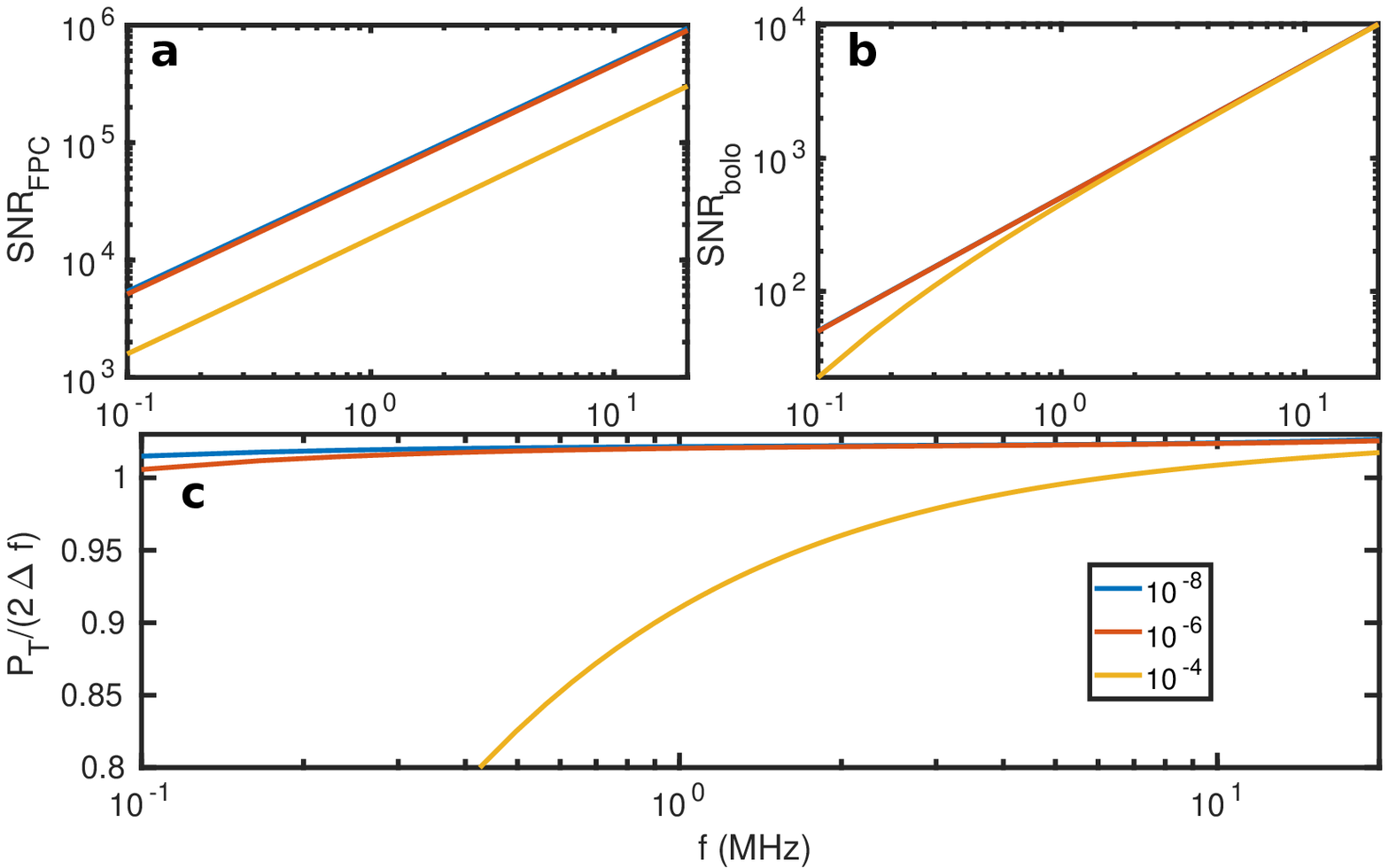}
\caption{\textbf{FPC and bolometer noise.} (a) FPC signal-to-noise ratio, the colors correspond to the legend in panel (c) and show the Dynes parameter. (b) Bolometer power detection signal-to-noise ratio, the blue curve is hidden behind the red one. (c) Total power produced by FPC.}
\label{f8}
\end{figure}

In Fig.~\ref{f8} Eqs.~\eqref{e17} and \eqref{e18} are depicted as function of the driving frequency for a device with the same parameters to the one simulated in Fig. 3f, but with $T_\mr{N}=100\,\mr{mK}$. This higher island temperature makes $P/(2\Delta f)>1$ and higher as shown in Fig.~\ref{f8}c. It is also evident that $\mr{SNR}_\mr{bolo}\ll \mr{SNR}_\mr{FPC}$, hence the bolometer detection noise is dominant over the FPC one at low frequencies (in fact, even when $f$ is large and $T\approx T_\mr{b}$ is not valid). While this holds independent of the Dynes parameter $\eta$ value (see Fig.~\ref{f8}), at low frequencies FPC $P$ is below Eq. (1) for devices with high $\eta$. For these structures the probability of an electron leaking into subgap states $\varepsilon <\Delta$ increases with decreasing driving frequency. Therefore, at low frequencies, when $\mr{SNR}_\mr{bolo}$ decreases, the main impediment for accurate FPC is having non-vanishing $\eta$.

\section{Calculation of $P_\mr{R}/P_\mr{L}$ at $V_\mr{b}=0\,\mr{V}$}

In general, Eqs. \eqref{e8} and \eqref{e13} can be expressed as
\begin{align}
\Gamma^\mr{L/R}_{n\rightarrow n\pm 1}&=\dfrac{1}{\rt}\xi^\mr{L/R}_{n\rightarrow n\pm 1}\left(\varepsilon^{\pm,\mr{L/R}}_{1e}\left(n,\vb\right),T_\mr{N},T_\mr{S}^\mr{L/R}\right)\\
\dot{Q}^\mr{N,L/R}_{n\rightarrow n\pm 1}&=\dfrac{1}{\rt}\dot{q}^\mr{N,L/R}_{n\rightarrow n\pm 1}\left(\varepsilon^{\pm,\mr{L/R}}_{1e}\left(n,\vb\right),T_\mr{N},T_\mr{S}^\mr{L/R}\right).
\end{align}
Where, $T_\mr{N}$ is the temperature of the normal-metal island and $T_\mr{S}^\mr{L/R}$ the quasiparticle temperature of the left/right lead. The quantities $\xi$ and $\dot{q}$ are functions of the island charge state and temperatures as well as bias voltage. When $\vb=0$ the energy change of the process becomes 
\begin{equation}
\varepsilon^{\pm,\mr{L}}_{1e}\left(n\right)=\varepsilon^{\pm,\mr{R}}_{1e}\left(n\right)=\varepsilon=\pm 2E_\mr{c}\left(n-n_\mr{g}\pm 0.5\right).
\end{equation}

Therefore, and making the important but reasonable assumption that $T_\mr{S}^\mr{L}=T_\mr{S}^\mr{R}$, we have that $\xi^\mr{L}_{n\rightarrow n\pm 1}=\xi^\mr{R}_{n\rightarrow n\pm 1}=\xi_{n\rightarrow n\pm 1}$ and $\dot{q}^\mr{N,L}_{n\rightarrow n\pm 1}=\dot{q}^\mr{N,R}_{n\rightarrow n\pm 1}=\dot{q}^\mr{N}_{n\rightarrow n\pm 1}$. As a consequence, the same reasoning can be done to $\dot{Q}_{n\rightarrow n\pm 1}^\mr{S,L/R}$, therefore the total power injected to the left/right superconductor is given by 
\begin{equation}
P_\mr{L/R}=\dfrac{1}{\rt}\sum_n{p_n\left(\dot{q}^\mr{S}_{n\rightarrow n+1}+\dot{q}^\mr{S}_{n\rightarrow n-1}\right)}.
\end{equation}
Where $\dot{q}^\mr{S}=\dot{q}^\mr{N}-\varepsilon\xi$. The ratio is then given by
\begin{equation}
\dfrac{P_\mr{R}}{P_\mr{L}}=\dfrac{1/R_\mr{R}\sum_n{p_n\left(\dot{q}^\mr{S}_{n\rightarrow n+1}+\dot{q}^\mr{S}_{n\rightarrow n-1}\right)}}{1/R_\mr{L}\sum_n{p_n\left(\dot{q}^\mr{S}_{n\rightarrow n+1}+\dot{q}^\mr{S}_{n\rightarrow n-1}\right)}}=\dfrac{R_\mr{L}}{R_\mr{R}}.
\end{equation}
This result is also true for the experiment on average as shown in Figs. 3g and 3h.

\section{Further data on $V_\mr{b}\neq 0\,\mr{V}$}

\subsection{Pumping against $\ag$}
\begin{figure}[ht!]
\includegraphics[scale=1]{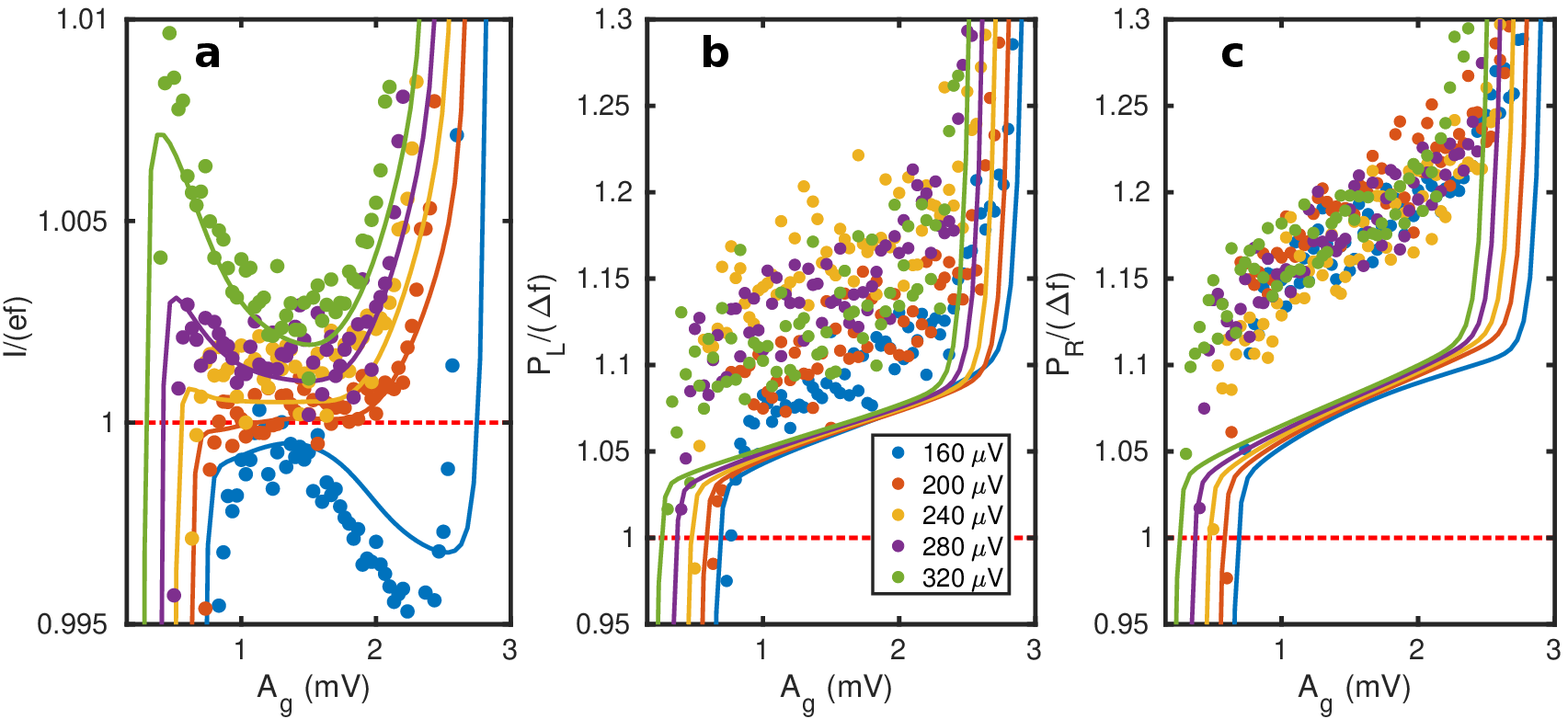}
\caption{\textbf{Current and injected power at $f=20\,\mr{MHz}$.} Extension of Fig. 4a. (a) Measured (dots) and simulated (lines) pumped current, colors correspond to the legend of panel (b). Data for $V_\mr{b}=240\,\mr{\mu V}$ has been included for completeness. (b) As in (a) for the power injected to the left lead. (c) As in (b) for the right lead.}
\label{f4}
\end{figure}

\begin{figure}[ht!]
\includegraphics[scale=1]{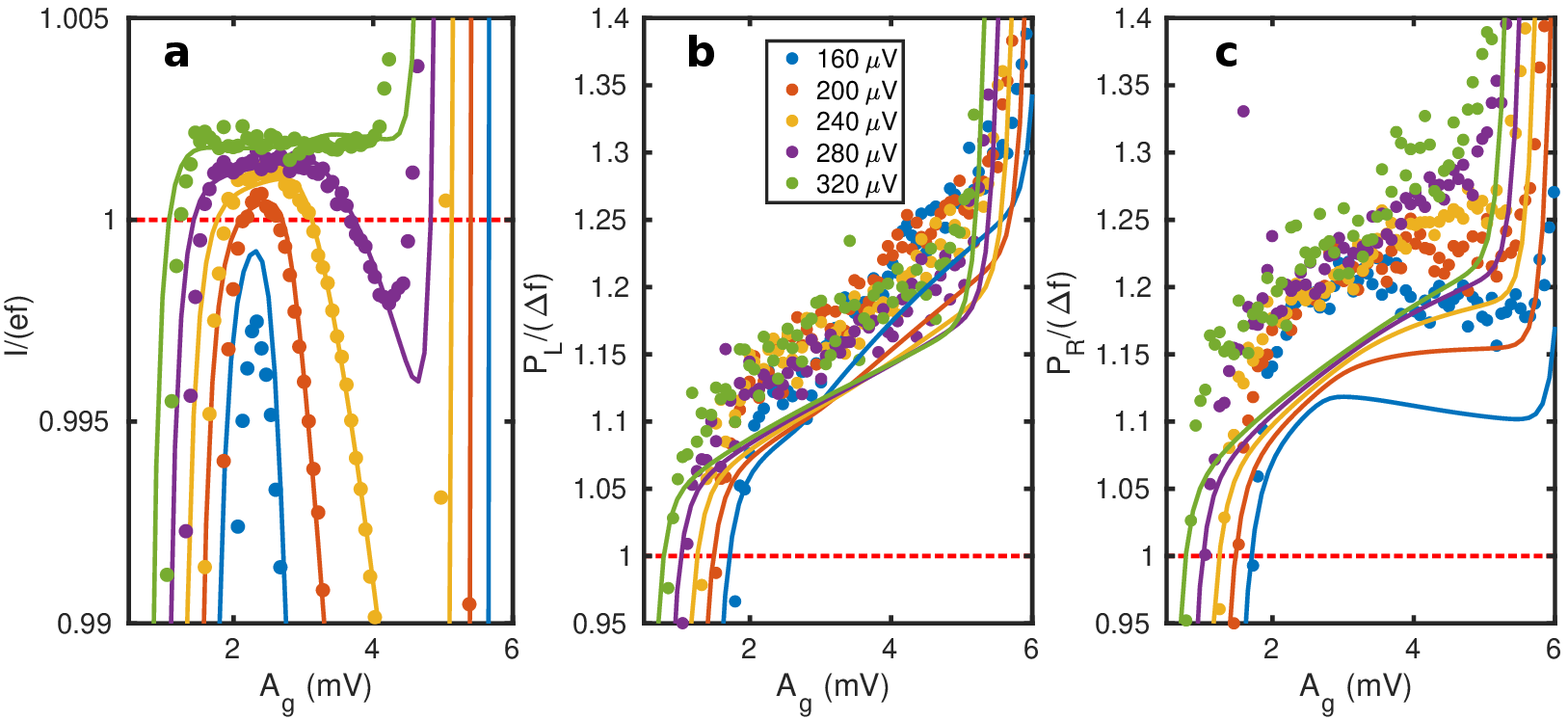}
\caption{\textbf{Current and injected power at $f=60\,\mr{MHz}$.} Extension of Fig. 4b. (a) Measured (dots) and simulated (lines) pumped current, colors correspond to the legend of panel (b). Data for $V_\mr{b}=160\,\mr{\mu V}$ has been included for completeness. (b) As in (a) for the power injected to the left lead. (c) As in (b) for the right lead.}
\label{f5}
\end{figure}

For the sake of completeness, we present further current and injected power data at different biases for driving frequencies $f=20\,\mr{MHz}$ and $f=60\,\mr{MHz}$, corresponding to Figs. 4a and 4b, respectively. These data are shown in Figs. \ref{f4} and \ref{f5}. Both sets of data and simulations follow the behavior described in the discussion pertaining to Fig. 4. On Fig. \ref{f5}b the injected power to the left lead decreases with higher biases for certain $\ag$ values. This behaviour is well explained by the fact that at higher biases energy injection is more equally distributed to both leads, therefore less power goes through the left junction while more is injected through the right one. This kind of behavior would also be visible at lower frequencies if the plateaus were measured at low enough biases as is clear from Fig. 4c.

\begin{figure}[ht!]
\includegraphics[scale=0.9]{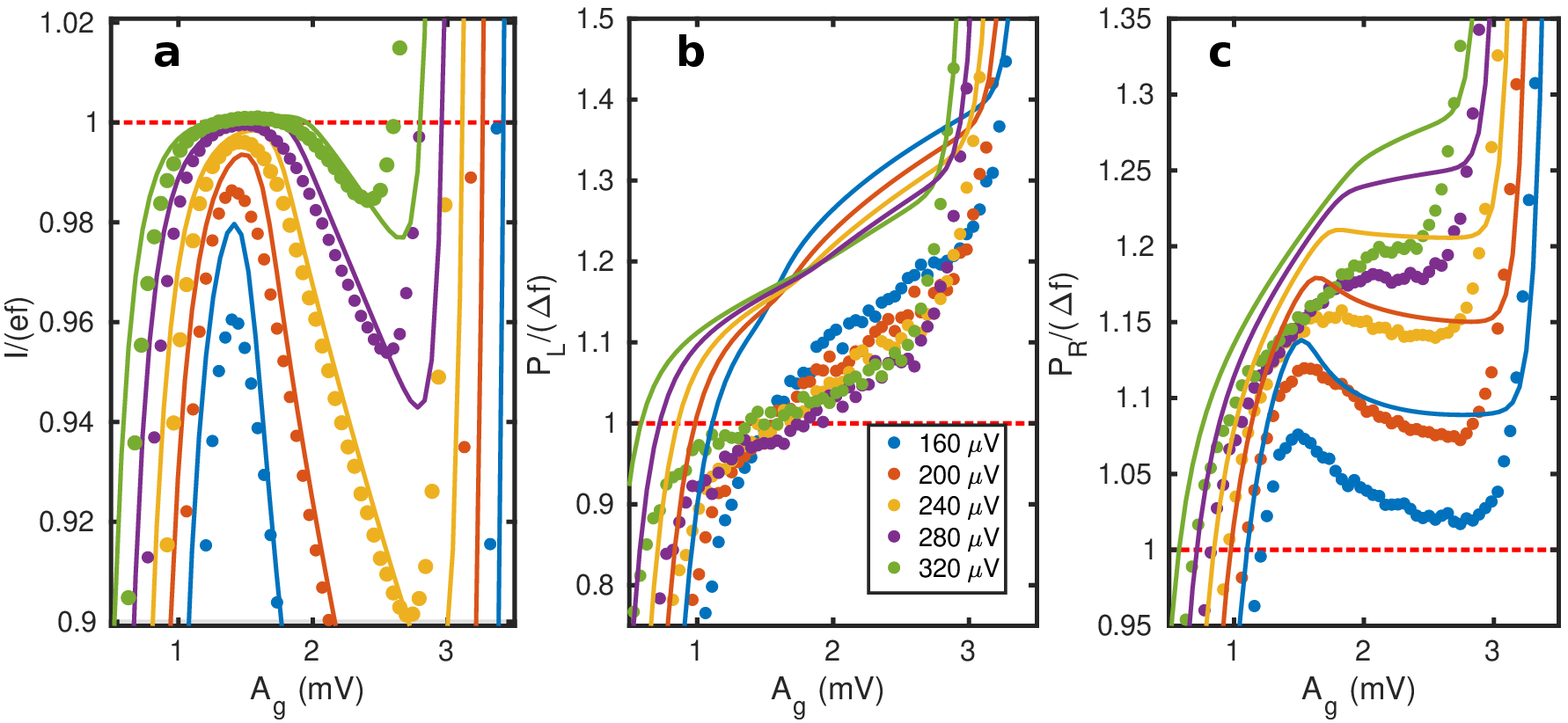}
\caption{\textbf{Current and injected power at $f=120\,\mr{MHz}$.} Data measured at $T_\mr{b}=130\,\mr{mK}$. (a) Measured (dots) and simulated (lines) pumped current, colors correspond to the legend of panel (b). (b) As in (a) for the power injected to the left lead. (c) As in (b) for the right lead.}
\label{f6}
\end{figure}

The previous measurements are complemented by Fig. \ref{f6}. These data were taken with a driving frequency of $f=120\,\mr{MHz}$. Again, the behavior of the power curves is in agreement with the description of Fig. 4. Notice how the plots in Fig. \ref{f6}b have the same pattern as those in Fig. \ref{f5}b. Since the driving frequency is higher this behaviour is more notorious in the former. The measured power dependency on $\ag$ follows the model although all the datasets are systematically lower. The higher bath temperature in which the measurements were done has reduced the quasiparticle trapping efficiency and therefore less heat is dissipated in the normal-metal bolometer. However, the amount of injected power through the transistor junctions is not expected to vary. We conclude that these errors are not due to the uncertainty in bolometric power detection.

\subsection{Pumped current against $V_\mr{b}$}

\begin{figure}[ht!]
\includegraphics[scale=1.0]{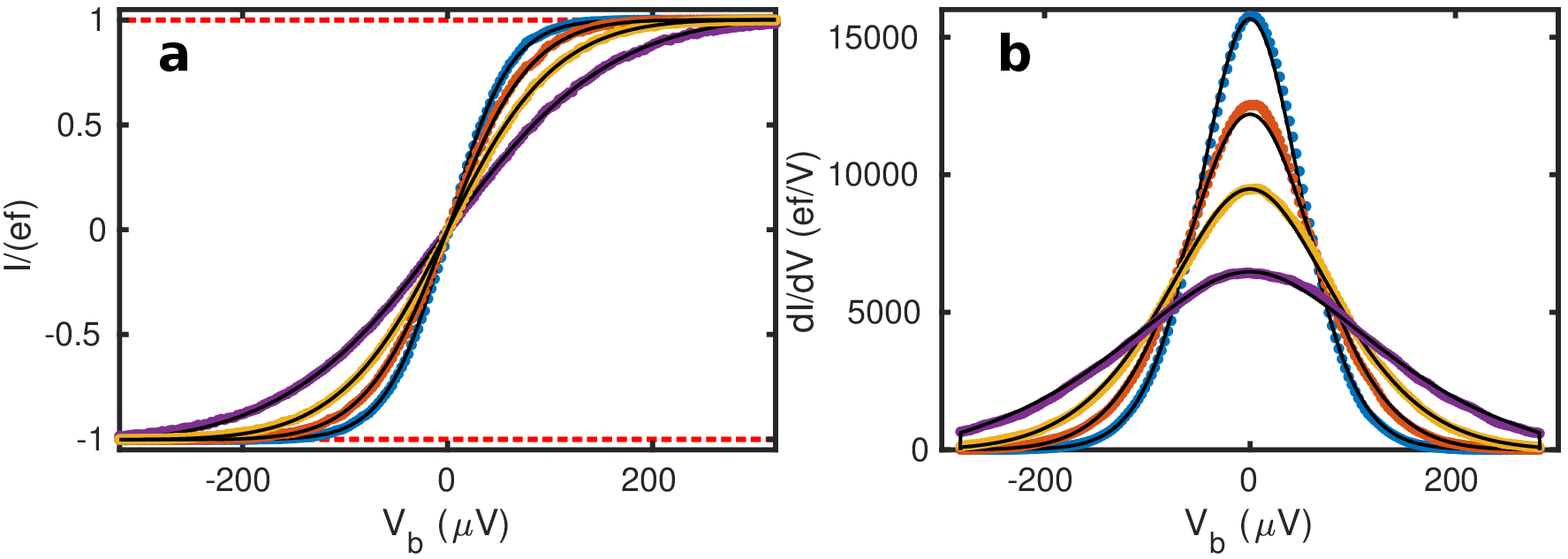}
\caption{\textbf{Pumped current against bias voltage.} (a) Experimental pumped current at constant amplitude with a driving frequency of $30,\,60,\,80,\,160\,\mr{MHz}$ as blue, red, yellow and purple dots, respectively. Black lines are calculations made with a Markovian equation, the dashed lines designate the ideal pumped current $I=ef$. (b) Differential conductance obtained by numerical differentiation of panel (a) data. Black lines are numerical derivatives of calculations of panel (a).}
\label{f7}
\end{figure}

We plot the pumped current against the bias at constant gate amplitude in Figure \ref{f7}. These measurements complement the results depicted in Figures 4c and 4d. Notice that the step-like shape of plots in Figure \ref{f7}a becomes sharper for frequencies not comparable to the transition rates of the device. This is more noticeable in the differential conductance (Figure \ref{f7}b, numerically calculated from data in panel a) where sharp peaks form around $V_\mr{b}=0$ for low frequencies, while at high frequencies these peaks are more spread and lower. This behavior is well described by the cartoons of Fig. 4c. Towards $V_\mr{b}=0$ the uncorrelated and non-unidirectional tunnelling events decrease the pumped current. While at high (absolute) bias values the unidirectional synchronized tunnelling events keep the current at $I\approx ef$. For high frequencies back-tunnelling events appear and the current decreases.

\section{Loss of energy selectivity at higher frequencies}

\begin{figure}[ht!]
\includegraphics[scale=0.8]{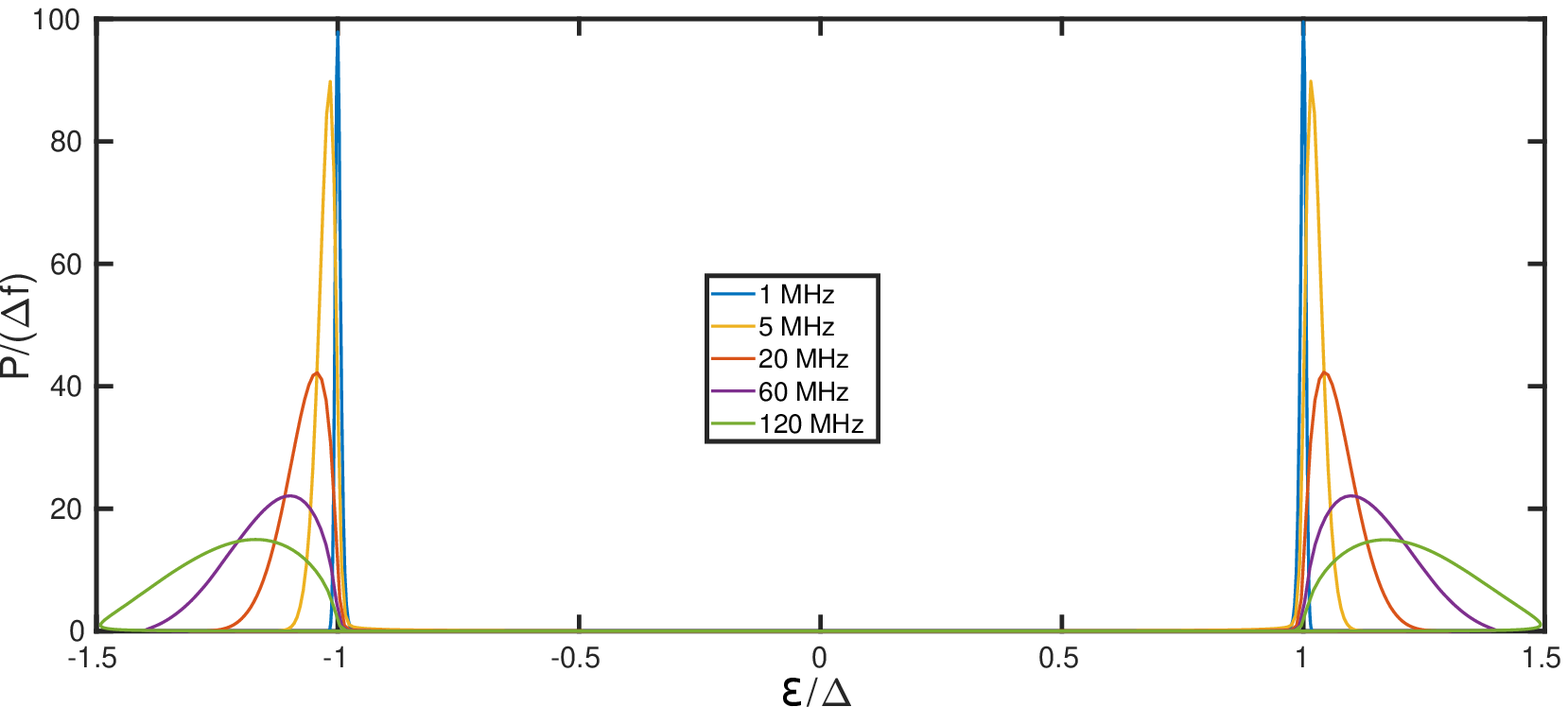}
\caption{\textbf{Change of the energy selectivity with driving frequency.} Instantaneous total injected power calculated under the same conditions as Fig. 3e but for different driving frequencies against the electron energy change when jumping into the island. Showing how electrons of higher energy can tunnel and selectivity is progressively lost the higher the driving rate.}
\label{f3}
\end{figure}

Recall from Section~\ref{S4} and Eq. 5 that the instantaneous power can be calculated as
\begin{equation}
P_\mr{R/L} =\sum_n{p\left(n\right)\left(\dot{Q}^\mr{S,R/L}_{n\rightarrow n+1}+\dot{Q}^\mr{S,R/L}_{n\rightarrow n-1}\right)},
\label{e19}
\end{equation}
and that the integrals $\dot{Q}^\mr{S,R/L}_{n\rightarrow n+1}$ depend on the energy change of the electron involved in the charging event, which in turn depends on time in the turnstile operation. As mentioned in the main manuscript for an electron tunnelling into the island, this energy is given explicitly by
\begin{equation}
\varepsilon = 2E_\mr{c}\left(0.5-n_\mr{g}\right),
\label{e20}
\end{equation}
when $V_\mr{b}=0$. Under sinusoidal gate driving $n_\mr{g}= 0.5+C_\mr{g}A_\mr{g}/e\sin{\left(2\pi ft\right)}$, and $t$ is now a parameter for $P=P_\mr{L}+P_\mr{R}$ and $\varepsilon$.

Figure \ref{f3} shows a parametric curve of the total power $P$ injected versus $\varepsilon$ under the same conditions as the results of Fig. 3e. Here, as well, it is possible to see that only electrons within a narrow energy band around the gap tunnel (inject energy) when the driving frequency is low and sharp power peaks around $\varepsilon=\Delta$ appear. As the driving frequency increases and the rate of change of $\varepsilon$ becomes comparable to the tunnelling rates, these peaks spread more in $\varepsilon$, indicating that electrons with higher energies are tunnelling and therefore increasing the average injected power.